\documentclass[reprint,amsmath,amssymb,aps,floatfix]{revtex4-2}

\usepackage[utf8]{inputenc}
\usepackage[colorlinks=true,linkcolor=blue,citecolor=blue,urlcolor=blue]{hyperref}

\usepackage{amsfonts}
\usepackage{mathtools}
\usepackage{amsthm}
\usepackage{bm}
\usepackage{bbm}
\usepackage{upgreek}
\usepackage{esint}
\usepackage[frak=esstix]{mathalpha}
\usepackage{slashed}

\usepackage{physics}
\usepackage{braket}
\usepackage{simpler-wick}

\usepackage{graphicx}
\usepackage[caption=false]{subfig}
\usepackage{booktabs}
\usepackage{dcolumn}
\usepackage{array}
\usepackage{algorithmicx}
\usepackage{algpseudocode}

\usepackage{tikz-feynman}

\usepackage[dvipsnames]{xcolor}
\usepackage{color,soul}
\usepackage{comment}
\usepackage[normalem]{ulem}
\usepackage{verbatim}
\usepackage{enumitem}

\usepackage{listings}
\setlist[itemize]{parsep=1em, itemsep=1em}
\setlist[enumerate]{parsep=0em, itemsep=1em}

\newcommand{\PreserveBackslash}[1]{\let\temp=\\#1\let\\=\temp}
\newcolumntype{C}[1]{>{\PreserveBackslash\centering}p{#1}}

\newcommand{\rom}[1]{\uppercase\expandafter{\romannumeral #1\relax}}

\newcommand{\EE}{\ensuremath{\mathcal{E}}}

\newcommand{\II}{\ensuremath{\mathcal{I}}}

\newcommand{\PP}{\ensuremath{\mathcal{P}}}

\newcommand{\al}{\ensuremath{\alpha}}

\newcommand{\de}{\ensuremath{\delta}}
\newcommand{\ep}{\ensuremath{\epsilon}}

\newcommand{\sg}{\ensuremath{\sigma}}

\newcommand{\De}{\ensuremath{\Delta}}

\newcommand{\pr}[1]{\left(#1\right)}

\newcommand{\Em}{\ensuremath{\mathbb{E}(m)}}

\begin{document}

\title{Optimizing resource bounds in direct fidelity estimation}

\author{Netanel Barel}
\email{mail: netanel.barel@quantum-art.tech}
\author{Lee Peleg}
\author{Yotam Kadish}
\author{Amit Ben Kish}
\author{Yotam Shapira}

\affiliation{Quantum Art, Ness Ziona 7403682, Israel}

\begin{abstract}
Direct fidelity estimation provides a way to estimate the fidelity between an experimentally prepared state and a desired pure target state without performing full tomography. Two influential formulations were introduced in 2011 by Flammia and Liu and by da Silva, Landon-Cardinal, and Poulin. In these protocols, the total estimation error is controlled through two distinct probabilistic steps: first, the fidelity is approximated using randomly sampled Pauli observables; second, each sampled expectation value is estimated from finitely many measurement outcomes. In this work we show that additional structural information about the noise can substantially sharpen the corresponding resource bounds. In particular, for some canonical channels the effective number of sampled Pauli settings can be reduced, leading to lower measurement cost both in the general pure-state setting and in the case of a stabilizer state. These results illustrate a broader point: worst-case confidence bounds in direct fidelity estimation can be significantly conservative when experimentally relevant structure is ignored. As a technical ingredient, we also revisit the allocation of the total accuracy and confidence budgets between the two probabilistic steps. Reformulating the analysis in terms of separate error parameters yields a constrained optimization problem whose solution lowers the average number of measurements in the general pure-state setting. Numerical simulations based on quantum circuits implemented in Qiskit illustrate both the improvement obtained under structured-noise assumptions and the conservativeness of the original worst-case bounds.
\end{abstract}

\maketitle


\section{Introduction}

Quantum state tomography is the standard tool to obtain a complete description of a prepared quantum state. Without any assumptions on its structure, the number of copies required by the protocol grows exponentially in the number of qubits, rendering it unrealistic beyond a few qubits \cite{Nielsen:2012yss}. In many situations, however, one does not seek a complete description of the output state $\sg$; rather, one wants a certificate that the prepared state is close to a prescribed pure target state $\rho$. Direct Fidelity Estimation (DFE) addresses precisely this task. The two 2011 works of Flammia and Liu \cite{FlammiaLiu2011} and of da Silva, Landon-Cardinal, and Poulin \cite{Silva:2011cbn} showed that fidelity can be estimated by measuring only a small subset of Pauli observables, sampled according to an importance distribution. These protocols have since become a standard point of reference in the broader literature on quantum-state certification \cite{Hangleiter2021}. In these works stabilizer states requires less measurements than the general pure-state, independent of the qubit's number, making the DFE protocol useful for them. We will treat these two cases separately along this work.

Subsequent works related to DFE developed the original two 2011 works. One early direction was the extension of the same sampling philosophy from states to quantum processes, as already emphasized in the original works, and implemented experimentally in \cite{Steffen:2012alj}. A second direction concerned the fundamental limitations of DFE-type methods: in particular, \cite{Leone:2022lrn} showed that for generic states the sample complexity of DFE is controlled by nonstabilizerness, and in general grows exponentially with the stabilizer R'enyi entropy. On the algorithmic side, later work sought to reduce measurement cost while preserving the Pauli-sampling spirit of the original protocol, for example by grouping commuting Pauli operators before sampling \cite{Barbera-Rodriguez:2025ouq}. More recent variants have also adapted the DFE viewpoint to special settings, such as stabilizer-state verification with few observables and many shots \cite{Theis:2024ofc} and Pauli-measurement shadow schemes inspired by DFE importance sampling \cite{Cha:2025wgn}.

The starting point of the present work is the observation that the standard resource bounds of direct fidelity estimation are worst-case bounds, and may therefore be substantially conservative when additional physical structure is available. Our main goal is to show that such structure can be incorporated into the analysis in a simple way, leading to significantly smaller measurement costs while keeping the original estimator and the overall logic of the protocol unchanged. In particular, we study noise channels of the form: 
\begin{equation}\label{eq: chap 1 channel description}
    \pr{1-p}\II + p\EE,    
\end{equation}
where $p<1$ is known (or can be bounded), and show that this assumption can markedly reduce both the number of relevant Pauli samples and the total number of measurements.\footnote{It is also worth keeping logically separate an additional idea that motivates part of our discussion. If independent information suggests that the prepared state is already close to the target, say with fidelity at least $1-p$, one may imagine a two-stage strategy in which a coarse estimate is first obtained and a sharper estimate is performed only when needed. The channel model $\pr{1-p}\II+p\EE$ is relevant to this intuition, but it is not implied by a fidelity lower bound alone. We therefore treat the structured-noise analysis as an additional independent assumption, not as a consequence of a bound of the form $F\ge 1-p$.} This class is physically meaningful in its own right: it includes standard examples such as depolarizing and dephasing-type channels, and it is closed under composition. More importantly for the present context, it provides a concrete illustration of a broader point: once physically motivated structure is taken into account, the usual confidence bounds of DFE can become significantly sharper.

A side theme of this work is a refinement of the baseline resource analysis itself. In the formulation of Flammia and Liu \cite{FlammiaLiu2011}, the total additive error is obtained by combining two probabilistic steps: one associated with the random sampling of Pauli observables, and the other associated with the finite estimation of the expectation value of each sampled observable. The same accuracy and confidence parameters are assigned to both steps. While this symmetric assignment is natural as a first presentation, it is typically not optimal in terms of the expected total number of measurements $\Em$ \footnote{In \cite{Silva:2011cbn}, where the total accuracy is written as a sum of two contributions. What is missing there, however, is an explicit optimization of the associated accuracy and confidence parameters with respect to the expected total number of measurements.}. Reformulating the argument with separate accuracy and confidence budgets for the two steps leads to a constrained optimization problem. In the general pure-state case, solving this problem decreases $\Em$; in the large qubits' number regime, the improvement includes a factor $4\ln\pr{4/\De}/\ln\pr{2/\De} \geq 4$ relative to the symmetric assignment, where $1-\De$ is the confidence level of the fidelity estimation procedure. For a stabilizer state, by contrast, we show that the original allocation is already optimal within the same framework. In the present work, this optimization plays two roles: it improves the baseline bound already in the general case, and it is essential for correctly exposing the savings that arise under additional structure assumptions. Without such a reallocation, one may even be led to the misleading conclusion that the required number of measurements increases when adding some additional structure.

Finally, we support the analytical discussion with numerical simulations carried out in Qiskit \cite{Qiskit2024}. These simulations illustrate both the gain obtained from the optimized budget split and, more significantly, the conservativeness of the original worst-case bounds when compared with the empirically observed distribution of the estimator.

The organization of this work is as follows. Section \ref{sec: prelim} recalls the DFE framework and fixes notation. \ref{sec: optimize} analyzes the optimized split. Section \ref{sec: noise} studies additional structured-noise assumptions, in particular channels of the form $\pr{1-p}\II+p\EE$. Section \ref{sec: implementation} presents numerical illustrations based on Qiskit simulations. Technical derivations, optimization details, and implementation notes are collected in the appendices.


\section{Preliminaries}
\label{sec: prelim}

Let $\rho$ be a desired pure $N$-qubit target state, let $\sg$ denote the state prepared in the laboratory, and let:
\begin{equation}\label{eq: chap 2 fidelity}
    F\pr{\rho,\sg} = \Tr\pr{\rho\sg}
\end{equation}
be the fidelity to be estimated. Following \cite{FlammiaLiu2011}, one samples Pauli observables $\PP_k$, i.e. Pauli strings over $N$ qubits, according to the probability distribution $P_k$ induced by their coefficient $\rho_k \equiv \Tr\pr{\rho\frac{\PP_k}{\sqrt{d}}}$ in $\rho$ ($d \equiv 2^N$):
\begin{equation}
    P_k = \rho_k^2,\ \rho = \sum_k\rho_k\frac{\PP_k}{\sqrt{d}}.
\end{equation}

The resulting estimator is built in two steps. The first is the \textit{sampling step} - a random average over sampled Pauli operators $k_1,\dots,k_\ell$ out of the $4^N$ possible Pauli operators composing $\rho$, where $\ell$ is determined by the error $\ep_1$ and the confidence level $1-\de_1$. The error is caused due to a partial evaluation of \eqref{eq: chap 2 fidelity}, assuming each expectation value is known exactly. The estimator for the fidelity is:
\begin{equation}
    Y \equiv \frac{1}{\ell}\sum_{i=1}^\ell X_i,\ X(k) \equiv \frac{\sg_k}{\rho_k},\ \braket{Y} = \braket{X} = F,  
\end{equation}
where $k$ is a random variable sampled according to $P_k$, $X(k)$ is a function of this random variable, and $X_i$ are $\ell$ copies of $X(k)$. Then, by Chebyshev's / Hoeffding's inequality for general / stabilizer state $\rho$:
\begin{equation}\label{eq: chap 2 first inequality}
    P(|F-Y|\geq \ep_1) \leq 1-\de_1,
\end{equation}
where in order to satisfy this inequality one sets $\ell = \left\lceil\frac{1}{\de_1\ep_1^2}\right\rceil$ for a general state and $\ell = \left\lceil\frac{2}{\ep_1^2}\ln\pr{\frac{2}{\de_1}}\right\rceil$ for a stabilizer state.

The second is the \textit{estimating step} - an empirical approximation of the expectation value associated with each sampled label. The error is caused due to shot noise. Each $X_i$ is estimated by $\tilde{X}_i = \frac{1}{\rho_{k_i}}\cdot\frac{1}{\sqrt{d}m_{k_i}}\sum_{j=1}^{m_{k_i}}A_{ij}$, where $A_{ij} = \pm 1$ is the outcome of measuring $\PP_{k_i}$, and $m_{k_i} = \left\lceil\frac{2}{d\rho_{k_i}^2\ell\ep_2^2}\ln\pr{\frac{2}{\de_2}}\right\rceil $ - the number of repetitions used to estimate the expectation value of the $i$th sampled Pauli observable. It is determined by the error $\ep_2$ and the confidence level $1-\de_2$, such that by Hoeffding's inequality the random variable $\tilde{Y} \equiv \frac{1}{\ell}\sum_{i=1}^\ell \tilde{X}_i$ satisfies:
\begin{equation}\label{eq: chap 2 second inequality}
    P(|Y-\tilde{Y}|\geq \ep_2) \leq 1-\de_2
\end{equation}  

Finally, the estimator $\tilde Y$ satisfies:
\begin{equation}
    P(|F-\tilde{Y}|\geq E) \leq 1-\De,
\end{equation}
where the total error $E$ and the confidence level $1-\De$ arise as a sum of two contributions:
\begin{equation}\label{eq: chap 2 error and conf constraints}
    E = \ep_1+\ep_2,\ \De = \de_1+\de_2.
\end{equation}
In Ref.~\cite{FlammiaLiu2011} $\ep_1 = \ep_1 \equiv \ep,\ \de_1 = \de_1 \equiv \de$. Some few remarks on the protocol are given in Appendix \ref{app: protocol remarks}. 


\section{Optimized splitting}
\label{sec: optimize}

We want to optimize the total number of single-copy measurements $m$:
\begin{equation}
    m \equiv \sum_{i=1}^{\ell} m_{k_i}.
\end{equation}
In the general case, $m$ is itself random because both the sampled labels and the measurement counts depend on the protocol. We therefore optimize its expectation $\Em$. In the stabilizer case, all $m_{k_i}$ are the same ($\rho_{k_i} = 1$) and $\Em = m = \ell m_k$ is determined. In both cases $\Em$ as a function of $\epsilon_{1,2}$ and $\delta_{1,2}$ can be minimized  under the constraints in Eq. \eqref{eq: chap 2 error and conf constraints}.

\subsection{General state}

We begin with the same estimator and the same proof strategy as in Ref.~\cite{FlammiaLiu2011}. The point is not to alter the protocol itself, but to optimize the assignment of the parameters entering its analysis in Eqs. \eqref{eq: chap 2 first inequality},\eqref{eq: chap 2 second inequality}. An upper bound on the mean number of measurements, up to inessential additive constants, is (Eq. (10) in \cite{FlammiaLiu2011}):
\begin{equation}\label{eq: chap 3 Em general schematic}
    \Em \lesssim \frac{1}{\ep_1^2\de_1}+\frac{2d}{\ep_2^2}\ln\pr{\frac{2}{\de_2}}.
\end{equation}
The first term is due to the random sampling of Pauli observables, and the second from the finite estimation of the expectation value of each sampled observable. Equation \eqref{eq: chap 3 Em general schematic} makes clear that the two stages scale differently in $\ep_i$ and $\de_i$. Therefore the symmetric assignment $\ep_1 = \ep_2 = E/2,\ \de_1 = \de_2 = \De/2$ is generally not optimal. Instead, one obtains a constrained optimization problem:
\begin{equation}\label{eq: chap 3 master optimization}
\begin{split}
    &\min_{\ep_1,\ep_2,\de_1,\de_2}
    \left\{\frac{1}{\ep_1^2\de_1}+\frac{2d}{\ep_2^2}\ln\pr{\frac{2}{\de_2}}\right\}
    \\
    &\text{subject to:} \ \ep_1 + \ep_2 = E,\ \de_1 + \de_2 = \De.
\end{split}    
\end{equation}
For fixed $\de_1$ and $\de_2$, the optimization over $\ep_1$ and $\ep_2$ can be carried out analytically, leading to a non-symmetric allocation whenever the coefficients of the $1/\ep_{1,2}^2$ differ. The remaining optimization over the confidence parameters may be performed numerically. In practice one performs everything numerically. The improvement is shown below; here we note that for $d = 2^N \gg 1$ the second term is dominant, and it is minimized with $\ep_2 \approx E,\ \de_2 \approx \De$, leading to $\Em \lesssim \frac{2d}{E^2}\ln\pr{\frac{2}{\De}}$. Compared to the symmetric allocation this is an improvement of factor $4\ln\pr{\frac{4}{\De}}/\ln\pr{\frac{2}{\De}}$.

\subsection{Stabilizer state}

For stabilizer target states, the number of sampled Paulis satisfy $\ell = \left\lceil\frac{2}{\ep_1^2}\ln\pr{\frac{2}{\de_1}}\right\rceil$ \footnote{The factor is not written in \cite{FlammiaLiu2011}, but it can be easily calculated and appears explicitly in (131) in \cite{Hangleiter2021}, with $\al = 1$.}. The probabilities $P_k = d^{-1}$ are constant, and hence $m_k = \left\lceil\frac{2}{\ell\ep_2^2}\ln\pr{\frac{2}{\de_2}}\right\rceil$ independent of the sampled Pauli. In order to minimize $m = \ell m_k$, it is beneficial to decrease $m_k$ as much as possible at the expense of $\ell$. Since $m_k \geq 1$, one sets $m_k = 1$, minimizing $\ep_2$ to be:
\begin{equation}
    \ep_2 = \frac{E\sqrt{\ln\pr{\frac{2}{\de_2}}}}{\sqrt{\ln\pr{\frac{2}{\de_2}}} + \sqrt{\ln\pr{\frac{2}{\de_1}}}}.
\end{equation} 
Plugging in $m = \ell$ one gets:
\begin{equation}
    m = \left\lceil\frac{2}{E^2}\pr{\ln\pr{\frac{2}{\de_1}} + \ln\pr{\frac{2}{\de_2}}}^2\right\rceil,
\end{equation}
which is minimized for $\de_1 = \de_2 = \De/2$, and therefore $\ep_1 = \ep_2 = E/2$. 

\section{Structured-noise assumptions}\label{sec: noise}

Assume that the actual state $\sg$ is produced from the target by a channel of the form $\pr{1-p}\II + p\EE$, where $\EE$ is a quantum channel. Then $\sg = (1-p)\rho + p\rho_{\EE}, \ \rho_{\EE} \equiv \EE\rho$ being a state by itself. Under this assumption, one can reduce $\ell$ by a factor of order $p^2$ for both a general and a stabilizer state, and this in turn decreases $\Em$.  

\begin{figure}[t]
    \includegraphics[width=0.23\textwidth,keepaspectratio,clip]{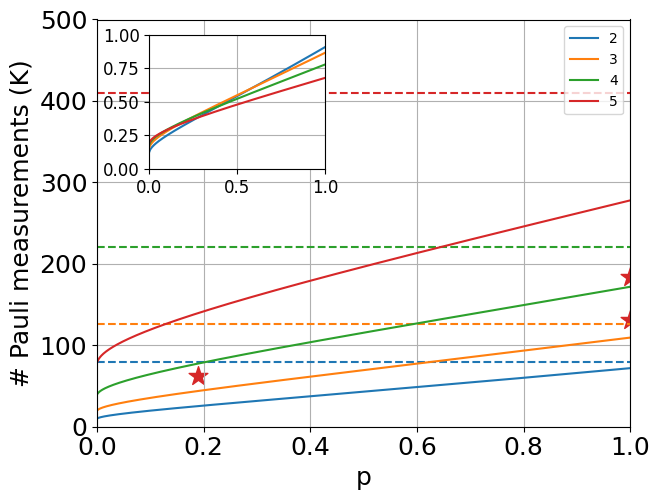}
    \includegraphics[width=0.23\textwidth,keepaspectratio,clip]{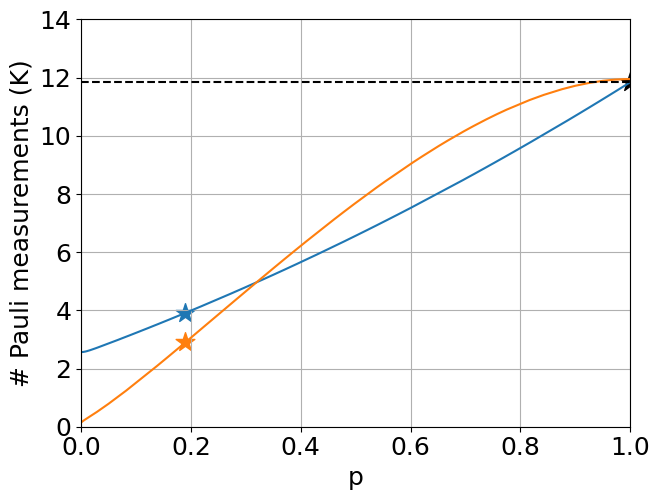}
    \caption{Left: upper bound on the mean number of measurements \eqref{eq: chap 3 Em general schematic} for a general state as a function of $p$ as described in \eqref{eq: chap 1 channel description}, for $E = 0.05$ and $\De = 0.1$, and for various qubit numbers (color). The dashed lines correspond to the upper bound with $\ep_1=\ep_2$ and $\de_1 = \de_2$ without using the additional information about the channel, their value grows with the qubits' number. The inset is the ratio of the solid lines to the dashed lines, marking the reduction in measurements afforded by our method. Right: number of measurements for a stabilizer state (it does not depend on the qubits' number, so there is only one curve for each qubits' number). The blue curve uses the additional information for the sampling step only, while the orange curve uses it for both steps, which is better for $p\leq 1/3$. Stars:  measurements' number sampled in the simulation presented in section \ref{sec: implementation} (exact for stabilizer, mean for general, with its standard deviation smaller than the marker's size).}
    \label{fig: chap 4 gain p2}
\end{figure}

For the general case, one modifies the sampling step and bounds $\text{Var}(X)$:
\begin{equation}
\begin{split}
    \text{Var}(X) &= \sum_k\sg_k^2 - \pr{\sum_k \sg_k\rho_k}^2 
    \\
    &= p^2\pr{\Tr\pr{\rho_{\EE}^2} - \Tr\pr{\rho\rho_{\EE}}^2} \leq p^2;
\end{split}
\end{equation}
the factor in the parenthesis being positive since it is the variance of the random variable similar to $X$ just with $\rho_{\EE}$ instead of $\sg$. Then Chebyshev's inequality is used:
\begin{equation}
\begin{split}
    P\pr{|Y - F|\geq \frac{\lambda p}{\sqrt{\ell}}} &\leq P\pr{|Y - F|\geq \lambda\sqrt{\text{Var}(Y)}} 
    \\
    & \leq \frac{1}{\lambda^2} \equiv \de_1.
\end{split}
\end{equation}
Choosing $\ep_1 \equiv \frac{\lambda p}{\sqrt{\ell}}$ leads to $\ell = \left\lceil\frac{p^2}{\ep_1^2\de_1}\right\rceil$. Then Eq. \eqref{eq: chap 3 master optimization} modifies with the first term multiplied by $p^2$. The corresponding improvement in the upper bound of $\Em$ is shown in Fig.~\ref{fig: chap 4 gain p2}.

For the stabilizer case, now not only $|X|\leq 1$ is bounded (so the sampling step was already modified compared to the general case, see \cite{FlammiaLiu2011,Silva:2011cbn}), but also its variance, so one uses one of Bernstein's inequalities to modify the sampling step:
\begin{equation}\label{eq: chap 4 first Bernstein}
    P\pr{\bigg|\frac{1}{\ell}\sum_{i=1}^\ell X_i - \braket{Y} \bigg| \geq \ep} \leq 2e^{-\frac{\ell \ep^2}{2\pr{p^2 + \frac{1}{3}\cdot 1\cdot\ep}}},
\end{equation}
leading to: 
\begin{equation}
    \ell = \frac{2\ln\pr{\frac{2}{\de_1}}}{\ep_1^2}\pr{p^2 + \frac{1}{3}\ep_1}.
\end{equation}
Following the same logic as above, one minimizes $\ep_2$ as to set $m_k = 1$:
\begin{equation}
    \ep_2 = \frac{E\sqrt{\ln\pr{\frac{2}{\de_2}}}}{\sqrt{\ln\pr{\frac{2}{\de_2}}} + \sqrt{p^2 + \frac{E-\ep_2}{3}}\sqrt{\ln\pr{\frac{2}{\de_1}}}}.
\end{equation}
This equation can be solved analytically in the two extreme cases where $p^2 \gg E$ or $p^2 \ll E$, and numerically for intermediate cases. Substituting into $m$ one should minimize:
\begin{equation}
    m = \frac{2}{E^2}\pr{\ln\pr{\frac{2}{\de_2}} + \sqrt{p^2 + \frac{\ep_1}{3}}\ln\pr{\frac{2}{\de_1}}}^2
\end{equation}   
to find $\de_2$. However, one can also bound $\text{Var}\pr{A^{\sg}_{ij}}$ and modify the estimating step as well. Only Paulis appearing in the decomposition of $\rho$ can be sampled; for these $|\Tr(\rho\PP) = 1|$, and so:
\begin{equation}
\begin{split}
    \text{Var}\pr{A_{ij}} & = p^2\text{Var}\pr{A^{\rho_{\EE}}_{ij}} + 2p(1-p)\pr{1-\Tr(\rho_{\EE}\PP)} 
    \\
    & \leq p^2\cdot 1 + 2p(1-p)(1+1) = p(4-3p).
\end{split}    
\end{equation} 
Using again Bernstein's inequality instead of Hoeffding's inequality for $\tilde{Y}$ gives $m_k = \left\lceil\frac{2}{\ell\ep_2^2}\ln\pr{\frac{2}{\de_2}}\pr{p(4-3p) + \frac{2\ep_2}{3}}\right\rceil$. As before $\ep_2$ is minimized as possible to set $m_k = 1$, and $\de_2$ is found to have minimal $m$. Using Bernstein's inequality helps only if the variance is small compared to the bound, which holds for $p\leq 1/3$. ($\ep_2$ is usually negligible because $E\ll 1$; the same is true for $\ep_1$ in \eqref{eq: chap 4 first Bernstein}, and there $p^2\leq 1$ so it always helps.) The improvement in the total number of measurements is shown in Fig. \ref{fig: chap 4 gain p2}. Indeed below $p=1/3$ it is beneficial to account for the structured noise in both the estimation and the sampling. (Although at $p=0$ the channel is perfect and measurement are not needed at all, as long as $p>0$ $m_k\geq 1$ and therefore the total number of measurements can not vanish.)

\section{Numerical illustrations and implementation}
\label{sec: implementation}

The decrease in the total number of measurements, besides of being important by itself, also sheds light on a more general phenomenon. As will be shown in this section, in numerical experiments, the histogram of the fidelity estimator may be much narrower than what the certified pair $\pr{E,\De}$ allows - i.e., that for large enough repetitions $M$ of the protocol, a much less than $M\De$ will be outside $(F-E,F+E)$ (if at all). Equivalently, values that the theorem treats as plausible outliers may in practice be essentially absent. This does not invalidate the theorem; it shows only that the theorem is a worst-case guarantee. The gap between the observed histogram and the proved interval is itself informative, because it indicates that unmodeled structure is suppressing fluctuations more strongly than the generic proof can capture.

To illustrate this, we performed quantum-circuit simulations in Qiskit \cite{Qiskit2024}. The target state is taken to be the operator $\exp\pr{-\frac{i}{2}\sum_{1 \leq n < m}^N\al_{nm}X_nX_m}$ acting on the all-up state $\ket{0}^{\otimes N}$, where $N$ is the number of qubits in the circuit. For general $\al_{nm}$ one obtains a general state, while for $\al_{nm}$ being integer multiples of $\pi/2$ the operator is Clifford, and so the resulting state is a stabilizer state. In the simulation we apply each gate $\exp\pr{-\frac{i}{2}\al_{nm}X_nX_m}$ separately, accompanied with a noise such that with probability $1-p_g$ nothing else happens, and with probability $p_g$ a reset error to $\ket{1}$ occurs for each of the participating two qubits (the noise is applied after the gate). For this type of noise we have $1-p = (1-p_g)^{2G}$, where $G$ is the number of gates applied (i.e., number of non-zero elements in the upper triangular part of $\al_{nm}$). In order to isolate the effect of the qubits' number from the noise model information we set $p=0.19$ (corresponding to $p_g=0.005$ for $N=7$). The remaining details of the simulation are given in appendix \ref{app: data}. 

\begin{figure}[t]
    \includegraphics[width=0.23\textwidth,keepaspectratio,clip]{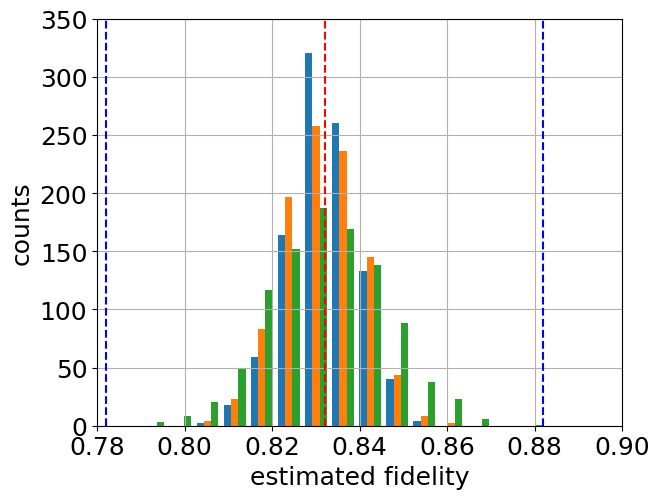}
    \includegraphics[width=0.23\textwidth,keepaspectratio,clip]{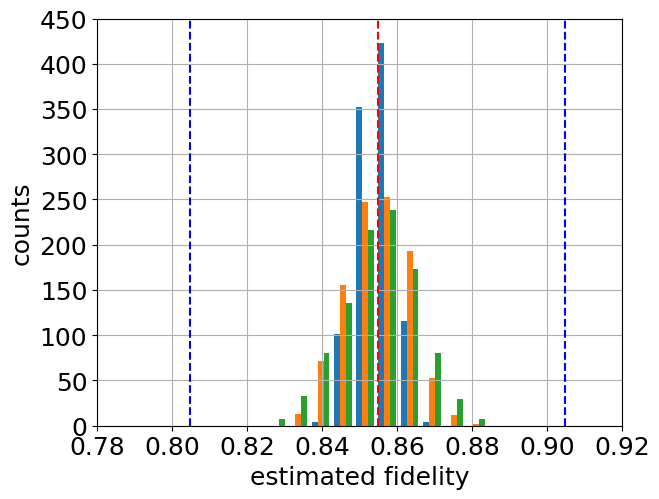}
    \caption{Histogram of 1000 fidelity estimations, for a general state (left) and for a stabilizer state (right), for $N=5$ qubits, $E=0.05$ and $\De = 0.1$. For the general state the symmetric allocation $\ep_1 = \ep_2,\ \de_1 = \de_2$ (blue, upper right star in the left panel of Fig.~\ref{fig: chap 4 gain p2}) is compared to the optimal allocation without using the information about the channel (orange, $p=1$, lower right red star), and to the optimal allocation when using this information (green, $p=0.19$, left red star). For the stabilizer state, the symmetric allocation (blue,black star in the right part of Fig.~\ref{fig: chap 4 gain p2}) - the optimal allocation without using the information about the channel - is compared to the optimal allocation modifying the sampling step alone (orange, $p=0.19$, blue star), and to the optimal allocation modifying both steps (green, $p=0.19$, orange star in Fig.~\ref{fig: chap 4 gain p2}).}
    \label{fig: chap 5 hist estimates}
\end{figure}

\begin{figure}[t]
    \includegraphics[width=0.23\textwidth,keepaspectratio,clip]{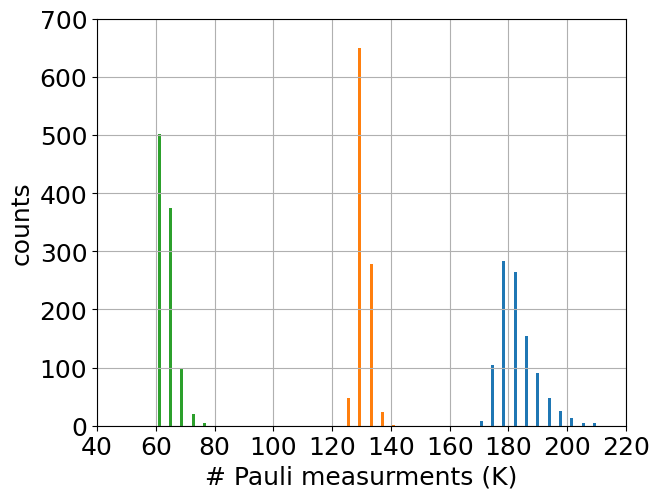}
    \includegraphics[width=0.23\textwidth,keepaspectratio,clip]{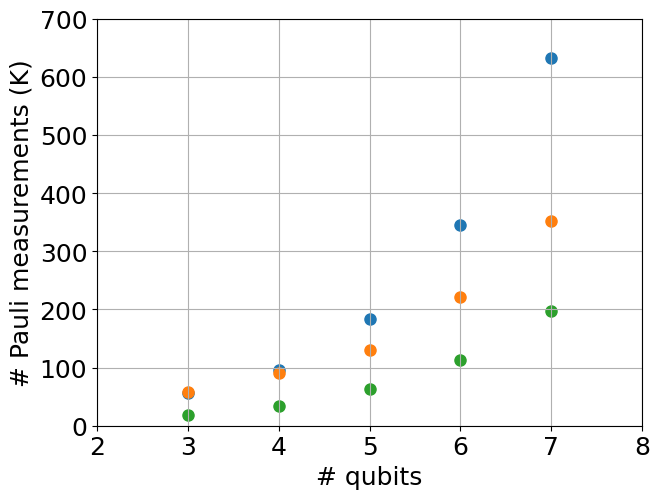}
    \caption{Left: histogram of total measurements' number used for a general state with $N=5$ qubits, with $E=0.05,\ \De=0.1$. The colors are the same as in the left part of Fig. \ref{fig: chap 5 hist estimates}. Right: the average number of measurements for several qubits' number (the std is smaller than circle size for all points).}
    \label{fig: chap 5 hist measurements}
\end{figure}

The results are shown in Fig.~\ref{fig: chap 5 hist estimates}, each histogram consists out of $M = 1000$ samples, for N = 5 qubits,
with $E = 0.05$ and $\De = 0.1$. Generally, broadening of the fidelity estimates means that for the same accuracy and confidence level guaranteed, one can perform less measurements. Equivalently, for the same amount of measurements, one can guarantee better accuracy / higher confidence level. One sees that the blue histograms, representing the protocol without using the knowledge about the noise channel, is very narrow compared to the allowed limits. For a stabilizer state (right panel) the sampling step alone broadens the histogram (orange), and it broadens further when modifying both steps (green). Still it does not exceed the $F\pm E$ bounds, and not even one sample out of the allowed $M\De = 100$ (on average, with a reasonable deviation of $\approx \sqrt{M\De} = 10$). For a general state (left part) the symmetric allocation is not optimal even when not using the information about the channel, so it broadens when using optimal allocation (orange), and further when using the information about the channel (green). For the stabilizer case the number of measurements decreased, while for the general case only its upper bound decreased, and therefore the broadening is less significant than for stabilizer state. Still the actual distribution is shifted towards smaller numbers, without any overlap between the distributions, as is shown in the left panel of Fig. \ref{fig: chap 5 hist measurements}. On the right panel the average number of measurements $E(m)$ is shown for the three protocol choice considered above, for states with increasing number of qubits.

\section{Conclusion and discussion}
\label{sec:conclusion}

We revisited the resource analysis of direct fidelity estimation. We showed that physically motivated assumptions can sharpen the bounds much more dramatically. For channels of the form $\pr{1-p}\II+p\EE$, the effective number of sampled Pauli observables and the total measurement cost can both decrease substantially. This decrement came with the observation that the two probabilistic steps in the standard analysis need not receive the same share of the total error and failure budgets for a prescribed accuracy and confidence level. Once these budgets are treated as optimization variables, the expected total number of measurements can be reduced in the general pure-state setting without changing the estimator itself.

More broadly, the results suggest that DFE admits a hierarchy of increasingly sharp resource bounds, depending on how much structure one is prepared to assume and justify. Even when one remains within the original protocol, careful optimization of the error allocation already improves the copy complexity. When one supplements the analysis with meaningful physical assumptions, the improvement may be much larger. Within the same accuracy and confidence level, one can use significantly less measurements and still remain within the allow fidelity bounds. We note that further research is warranted as the histogram of estimated fidelities still does not reach the bounds. It might be that the probabilistic inequalities used for deriving the protocol are not saturated from the beginning, or that there is more information to exploit and reduce the measurements' number further.


\bibliography{refs}


\appendix


\section{Joint probability space and the union bound}
\label{app: protocol remarks}

In this appendix we give couple of remarks about the DFE protocol. The first regards the role of $Y$ in the two steps presented in . We emphasize that $Y$, being a random variable in the first step whose mean equals $F$, is a constant in the second step, which is estimated by the random variable $\tilde Y$. It is useful to make the joint sample space explicit. Let
\begin{equation}
    K = \{k_1,\dots,k_{4^N}\}
\end{equation}
denote the random Pauli labels, and let
\begin{equation}
    A = \{A_{ij}\}
\end{equation}
collect all single-shot outcomes used to estimate the sampled expectation values. The subset of $A$ appearing in $\tilde{Y}$ depend on the labels sampled from $K$. Then the intermediate random variable $Y = Y(K)$ depends only on $K$, whereas the final estimator $\tilde{Y} = \tilde{Y}(A|K)$ depends on both $K$ and $A$. The finite-sampling inequality \eqref{eq: chap 2 second inequality} is naturally understood as a conditional statement given $K$; after averaging over $K$, it becomes an unconditional bound on the same joint probability space. This viewpoint justifies the final union bound while making transparent that the two stages correspond to distinct sources of randomness.

The next remarks regards the protocol itself. The identity operator is inside any state, and there is no need to sample it. Its percentage is about $2^{-N}$. One can modify the protocol such that it will not participate in the sampled Pauli operators at all. 

Additionally, for small number of qubits, a small number of Pauli operator compose the state, and they will be sampled many times. Since $m_k$ depends only of the Pauli $\PP_k$ sampled, one can measure the all appearances of $\PP_k$ together. We denote it by $m_{\PP}$, and by $n_{\PP}$ the number of occurrences of $\PP$ in the $\ell$ samples, satisfying:
\begin{equation}
    \ell = \sum_{\PP}n_{\PP}.
\end{equation}
Then trivial algebra shows:
\begin{equation}
\begin{split}
    \tilde Y = \frac{1}{\sum_{\PP}n_{\PP}}\sum_{\PP}\frac{1}{\tr\pr{\rho \PP}}\frac{1}{m_{\PP}}\sum_{j=1}^{n_{\PP}m_{\PP}}A_{\PP j}.
\end{split}
\end{equation}
The probabilities are reflected only through the $n_{\PP}$'s, and each Pauli is measured at once $n_{\PP}m_{\PP}$ times. Note that this is not some estimation on each trace which would be:
\begin{equation}
    \sum_{\PP}\frac{1}{\tr\pr{\rho \PP}}\frac{1}{n_{\PP}m_{\PP}}\sum_{j=1}^{n_{\PP}m_{\PP}}A_{{\PP}j}
\end{equation}  
where each Pauli is averaged alone.  

Last, the protocol can estimate the fidelity being larger than $1$. It should be possible to take this constraint into account, but for practical purposes one expects to choose an error such that $F+E < 1$. The same applies for the other bound $F>0$. 


\section{Simulation data}\label{app: data}

We took $N=5$ qubits. The matrices $\al_{nm}$ for the general case where randomized, with values:
\begin{equation*}
    \al^{gen}_{nm} = 
    \begin{pmatrix}
        * & 0.31 & 1.05 & 1.48 & 1.43
        \\
        * & * & -1.46 & -1.49 & -0.73
        \\
        * & * & * & 0.89 & -0.62
        \\
        * & * & * & * & 0.61
        \\
        * & * & * & * & *
    \end{pmatrix},
\end{equation*}
while for the stabilizer case $\al^{sta}_{nm} = -\pi/2$ for $(n,m) \in \{(1,2),(1,3),(1,4),(2,5),(3,4),(4,5)\}$ and vanishes for the other elements in the upper right triangle.

\end{document}